\documentclass[10pt,aps,pra,twocolumn,showpacs,floatfix,superscriptaddress]{revtex4-1}
\usepackage{graphicx}
\usepackage{epstopdf}
\usepackage[colorlinks]{hyperref}
\hypersetup{
 colorlinks=true,
 citecolor=blue,
 linkcolor=black,
 urlcolor=blue}
\usepackage{amssymb}
\usepackage{times}
\usepackage{amsmath,mathrsfs}
\usepackage{dsfont}
\usepackage{amsfonts}%
\newcommand{\bra}[1]{\langle#1|}
\newcommand{\ket}[1]{|#1\rangle}

\renewcommand{\sec}[1]{\hyperref[sec:#1]{Sec.~\ref{sec:#1}}}
\newcommand{\app}[1]{\hyperref[app:#1]{Appendix~\ref{app:#1}}}
\newcommand{\eq}[1]{Eq. (\ref{eq:#1})}
\newcommand{\fig}[1]{\hyperref[fig:#1]{Fig.~\ref{fig:#1}}}

\newcommand{\bfH}{{\hat{\mathcal H}}}
\renewcommand{\vec}[1]{{\bf #1}}

\begin{document}
\title{Linear Mode-Mixing of Phonons with Trapped Ions}
\author{Kevin Marshall}
\affiliation{Department of Physics, University of Toronto, Toronto, M5S 1A7, Canada}
\email{marshall@physics.utoronto.ca}
\author{Daniel F.V. James}
\affiliation{Department of Physics, University of Toronto, Toronto, M5S 1A7, Canada}

\date{\today}

\begin{abstract}
We propose a method to manipulate the normal modes in a chain of trapped ions using only two lasers.  Linear chains of trapped ions have proven experimentally to be highly controllable quantum systems with a variety of refined techniques for preparation, evolution, and readout, however, typically for quantum information processing applications people have been interested in using the internal levels of the ions as the computational basis.  We analyse the case where the motional degrees of freedom of the ions is the quantum system of interest, and where the internal levels are leveraged to facilitate interactions.  In particular, we focus on an analysis of mode-mixing of phonons in different normal modes to mimic the quantum optical equivalent of a beam splitter.
\end{abstract}
\maketitle

\section{Introduction}
The field of continuous-variable quantum information \cite{braunstein05}, where one considers a system of quantum harmonic oscillators, is largely framed within the context of quantum optics.  In this framework, the different modes of the electromagnetic field comprise the harmonic oscillators.  However, this is not the only framework one can consider; the motional quanta of mechanical systems, for example,  a linear chain of trapped ions gives rise to an equally valid set of harmonic oscillators.  Working with trapped ions can prove beneficial as one has a high degree of control \cite{haffner08}.  However, familiar elements from quantum optics, such as a beam splitter, are not necessarily straightforward to realize with phonons in a trapped ion system \cite{klau12}.  The first trapped ion quantum computing scheme, proposed by Cirac and Zoller \cite{cirac95}, utilizes only the center-of-mass phonon mode and only to mediate an interaction between spatially separated ions.  In this framework one wishes to cool \cite{diedrich89,monroe95} all motional modes to their ground state to avoid decoherence in the computational basis \cite{wineland98,djames99-heat,marquet03}, and as such the extra degrees of freedom provided by the normal modes of phonons are considered largely a nuisance \cite{cirac95}.

The task of building a robust quantum computer, capable of universal fault-tolerant computation, remains a daunting task to this date \cite{divincenzo05}.  In light of this, and with motivation to demonstrate the power of quantum computation over classical computers, \emph{post-classical computation} has attracted much interest in recent years \cite{aaronson11,broome13,spring13}.  The goal of this field is to look towards quantum systems which we can control effectively and to find problems well-suited to these systems which display a quantum-advantage, as oppose to choosing a problem and then finding a quantum system capable of implementing it.  A key example of such a problem was devised by Aaronson and Arkhipov \cite{aaronson11}, called boson sampling, and requires only linear mode-mixing between harmonic oscillators as well as the ability to count excitations in the readout stage.  Typically, this problem is considered in the framework of quantum optics where each mode is a spatial mode of the electromagnetic field, and the mixing is provided by an photons passing through an array of beam splitters.  However, previous work has studied the possibility of instead using phonons in trapped ion systems \cite{klau12,duan14,villar16} and this has been demonstrated in an interference experiment \cite{toyoda15}.

In this work we study a novel approach to implement linear mode-mixing between phonons in the normal modes of a linear chain of trapped ions.  In particular, we demonstrate the ability to generate two-mode Gaussian operations on the motional modes; non-Gaussian operations have also been investigated \cite{nie09}.  This is motivated both by the desire to study continuous-variable quantum operations embedded in the framework of phonons as well as to provide a primitive which may prove useful in post-classical quantum computations developed for trapped ion systems.  The paper is structured in the following way.  In \sec{interaction} we discuss the interaction that we will make use of in the main protocol, namely the interaction of a laser with one ion in the chain.  The effective Hamiltonian framework is briefly detailed as well as the resulting effective form for our interaction in \sec{effective}.  We use this framework to derive the specific case of a beam splitter in \sec{bsplitexample}.  We discuss the full protocol in \sec{protocol} and  justify neglecting off-resonant terms in \sec{resonant}.  The limitations of our scheme are presented in \sec{limitations} and we briefly discuss an example of phonon bunching, akin to the Hong-Ou-Mandel experiment, in \sec{HOM}.  Finally, we review the benefits of our approach in \sec{conclusions}.
\section{Interaction}\label{sec:interaction}
\begin{figure}[htp]
\centering
\includegraphics[scale=.6]{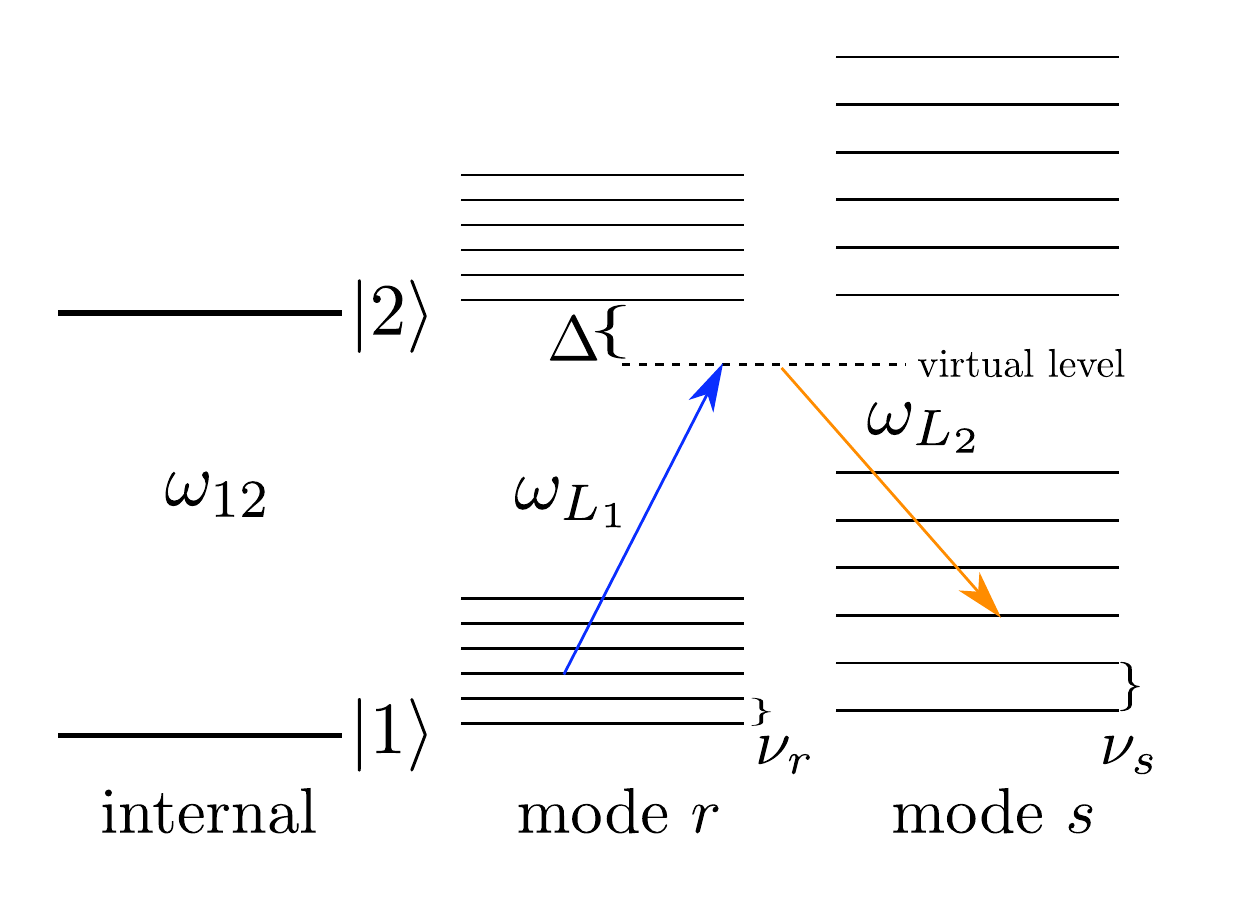}
\caption{(color online) Two lasers of frequency $\omega_{L_1}$ and $\omega_{L_2}$, both far detuned from the atomic transition frequency $\omega_{12}$, illuminate one ion in a linear chain.  A second-order interaction between normal modes $r$ and $s$ is resonantly excited through a virtual level. }
\label{fig:levels}\end{figure}

In the Cirac and Zoller scheme \cite{cirac95}, one uses phonons to mediate an interaction between the internal levels of two trapped ions.  Here, we develop a method similar but converse in nature where we mediate an interaction between phonons through the internal states of the ions with bichromatic light, more analogous to the M\o{}lmer-S\o{}rensen gate for qubits \cite{sorensen00}.  Consider a linear chain of $N$ trapped ions, each with mass $M$, where we restrict ourselves to consider only motion in the axial direction of the trap.  We use illuminate one ion in the chain with two lasers, of frequencies $\omega_{L1}$ and $\omega_{L2}$. These pulses are well-modelled by plane waves over some finite duration of interaction.  As indicated by the level diagram in \fig{levels}, we engineer an interaction between normal modes $r$ and $s$ by satisfying the resonance condition $\omega_{L_1}-\omega_{L_2}=\nu_s-\nu_r$.

The interaction Hamiltonian is then given by $\hat H=-\vec{E}(\hat{\vec{r}},t)\cdot \hat{\vec{d}}$, where we have left off the implicit time dependence of the operators $\hat{\vec r}$ and $\hat{\vec d}$.  Furthermore, all quantities will depend on which ion in the chain we use but we leave this dependence implicit to avoid cumbersome labels.  We make the two-level approximation for the internal levels of the ion, justified by the assumption that all other levels are far off resonance.  This lets us express the dipole operator as 
\begin{align}\label{eq:dipole}
\hat{\vec d}&=\vec{d}_{12}\ket 1\bra 2 e^{-i\omega_{12}t}+\vec{d}_{21}\ket 2\bra 1 e^{i\omega_{12}t}.
\end{align}
Making the plane wave approximation for the lasers, we can express the electric field as
\begin{align}\label{eq:field}
\vec{E}(\hat{\vec r},t)&=\vec{E}_1\cos\left(\vec{k}_{1} \cdot \hat{\vec r}-\omega_{L_1}t+\phi_{1}\right)\nonumber\\
&+\vec{E}_2\cos\left(\vec{k}_{2} \cdot \hat{\vec r}-\omega_{L_2}t+\phi_{2}\right),
\end{align}
where $\phi_{1,2}$ is the phase of the corresponding laser.  Finally, the ion position operator can be expressed as a small displacement from some equilibrium position $\vec{r}_0$
\begin{align}
\hat{\vec r}&=\vec r_0+\vec{e}_{\text{trap}}\hat q\nonumber\\
&=\vec r_0+\vec{e}_{\text{trap}}\sqrt{\frac{\hbar}{2NM\nu_{\text{trap}}}}\sum_{p=1}^N s^{(p)}\left(\hat a_p^\dagger e^{i\nu_p t}+\hat a_p e^{-i\nu_pt}\right),
\end{align}
where the sum runs over the normal modes, with frequencies $\nu_p$, and where $s^{(p)}$ are dimensionless ion-mode coupling constants defined as $s^{(p)}=\sqrt{N\nu_{\text{trap}}/\nu_p} b_\ell^{(p)}$ where $b_\ell^{(p)}$ are the eigenvectors defining the normal modes \cite{djames98}.

We can neglect off-resonant terms by making the rotating wave approximation when substituting \eq{dipole} and \eq{field} into the Hamiltonian.  This yields the following result
\begin{align}
\hat H_i&=-\frac{\vec E_i\cdot \vec d_{12}}{2}\ket 1\bra 2 e^{i(\omega_{L_i}-\omega_{12})t}\exp\left\lbrace-i\left(\vec k_{L_i}\cdot\hat{\vec r}+\phi_{i}\right)\right\rbrace\nonumber\\
&+h.c.
\end{align}
where $\hat H=\hat H_1+\hat H_2$.  The second term in the above expression can be expanded as
\begin{align}
&\exp\left\lbrace-i\left(\vec k_{L_i}\cdot\hat{\vec r}+\phi_{i}\right)\right\rbrace=\exp\left\lbrace-i\left(\vec k_{L_i}\cdot \vec r_0+\phi_{i}\right)\right\rbrace\times\nonumber\\
&\times\exp\left\lbrace-i\eta_i\sum_{p=1}^Ns^{(p)}\left(\hat a_p^\dagger e^{i\nu_pt}+\hat a_p e^{-i\nu_pt}\right)\right\rbrace,
\end{align}
where $\eta_i=\sqrt{\hbar k_{L_i}^2\cos^2\theta_i/2NM\nu_{\text{trap}}}$ is the Lamb-Dicke parameter with $\theta_i$ defined as the angle between the axis of the trap and the propagation direction of the $i$-th laser.  We to simplify the expression, we combine the phases to define $\psi_i=\vec k_{L_i}\cdot \vec r_0+\phi_{i}+\pi/2+ \text{Arg}(\vec E_i\cdot \vec d_{12})$, and the Rabi frequency is then defined as $\Omega_i=\left|\vec E_i\cdot \vec d_{12}\right|/\hbar$.

Under the Lamb-Dicke approximation we may expand the exponential as
\begin{align}
&\exp\left\lbrace-i\eta_i\sum_{p=1}^Ns^{(p)}\left(\hat a_p^\dagger e^{i\nu_pt}+\hat a_p e^{-i\nu_pt}\right)\right\rbrace\nonumber\\
&\approx \hat{\mathcal I}-i\eta_i\sum_{p=1}^Ns^{(p)}\left(\hat a_p^\dagger e^{i\nu_pt}+\hat a_p e^{-i\nu_pt}\right)
\end{align}
We arrive at the following expression for the interaction of laser ($i$) with ion ($\ell$).
\begin{align}\label{eq:Hfull}
&\hat H_i^{(\ell)}=\frac{\hbar \Omega_i}{2}e^{-i\psi_i}\hat \sigma_-^{(\ell)} e^{i\Delta_i t}\nonumber\\
&-\frac{i\hbar \eta_i\Omega_i}{2}e^{-i\psi_i}\sigma_-^{(\ell)}e^{i\Delta_i t}\sum_{p=1}^Ns^{(p)}\left(\hat a_p^\dagger e^{i\nu_pt}+\hat a_p e^{-i\nu_pt}\right)\nonumber\\
&+ h.c.,
\end{align}
where the detuning is defined as $\Delta_i=\omega_{L_i}-\omega_{21}$ and the atomic transition operator for the $\ell$-th ion is defined as $\hat\sigma_-^{(\ell)}=\ket 1\bra 2_\ell$ .  To construct a two-mode gate between distinct modes $r$ and $s$ we turn on the interaction $\hat H=\hat H_r^{(\ell)}+\hat H_s^{(\ell)}$, where we allow the two lasers to be of different frequencies in general.   Suppose we could implement a Hamiltonian of the form \eq{Hfull} with only one mode, we could then write this Hamiltonian in the form
\begin{align}\label{eq:Hharmonic}
\hat H&=\sum_{n=1}^4 \hat h_n^{(\ell)} e^{-i \omega_n t}+h.c.
\end{align}
by choosing
\begin{align}\label{eq:hn}
\hat h_1^{(\ell)}&=i\hbar G_1\sqrt{\frac{\hbar}{2M\nu_r}}\hat a_r\hat \sigma_-^{(\ell)}  &&\omega_1= \hphantom{-}\nu_r-\Delta_1\nonumber\\
\hat h_2^{(\ell)}&=i\hbar G_1\sqrt{\frac{\hbar}{2M\nu_r}}\hat a_r^\dagger\hat \sigma_-^{(\ell)}  &&\omega_2= -\nu_r-\Delta_1\nonumber\\
\hat h_3^{(\ell)}&=i\hbar G_2\sqrt{\frac{\hbar}{2M\nu_s}}\hat a_s\hat \sigma_-^{(\ell)}  &&\omega_3= \hphantom{-}\nu_{s}-\Delta_2\nonumber\\
\hat h_4^{(\ell)}&=i\hbar G_2\sqrt{\frac{\hbar}{2M\nu_s}}\hat a_s^\dagger\hat \sigma_-^{(\ell)}  &&\omega_4= -\nu_{s}-\Delta_2,
\end{align}
where $G_i=\Omega_i k_{L_i}\cos\theta_i e^{-i\psi_i}/2$ is a coupling strength, and where we have assumed that all choices of detunings are such that $\Delta_1\pm\Delta_2 \in \Omega( \nu_{\text{trap}})$ so that the first term in \eq{Hfull} can be safely neglected; up to a global phase shift.  Since the frequency of the atomic transition will dominate all other frequencies in the problem for typical ion traps, this choice of $\hat h_n^{(\ell)}$'s ensures that the sum of any two $\omega_j$'s will be large compared to the difference of any two.  For now, we keep all terms to be rigorous, however in the following section we will argue that under certain resonance conditions many terms can be neglected.
\section{Effective Hamiltonian}\label{sec:effective}
The full dynamics of the interaction specified by \eq{Hharmonic} would be difficult to study, and the situation would become even more dire once we considered transverse modes and the true interaction as in \sec{protocol}.  Fortunately, if we are only interested in sufficiently low frequencies then we can make use of the effective Hamiltonian formalism \cite{djames07,ogamel10}.  The effective Hamiltonian model provides a simple framework to study the effects of harmonic Hamiltonians, which contain high-frequency components, when restricted to finite time-resolution.

The effective Hamiltonian for \eq{Hharmonic} is given by
\begin{align}\label{eq:Heff}
\hat H_{eff}^{(\ell)} &=\sum_{n,m}^4\frac{1}{\hbar \omega_{nm}^+}\left[\hat h_m^{(\ell)\dagger},\hat h_n^{(\ell)}\right]e^{i(\omega_m-\omega_n)t},
\end{align}
where
\begin{align}\label{eq:wnm}
\frac{1}{\omega_{nm}^+}&=\frac{1}{2}\left(\frac{1}{\omega_n}+\frac{1}{\omega_m}\right).
\end{align}
In deriving this form we have assumed that terms which oscillate at the sum of any two $\omega$'s can be neglected; this condition is satisfied in \eq{hn} since $(-\Delta_i)\gg \nu_j~\forall i,j$.  Defining a matrix $\bfH^{(\ell)}$ such that $\bfH_{nm}^{(\ell)}$ corresponds to the term in the summation of \eq{Heff} containing $[\hat h_m^{(\ell)\dagger},\hat h_n^{(\ell)}]$, we can express the effective Hamiltonian as $\hat H_{eff}^{(\ell)}=\sum_{n,m}^4 \bfH^{(\ell)}_{nm}$.  The resulting form of this matrix for the terms in \eq{hn} can be found explicitly in \app{Hterms}.  In addition to the evolution governed by the effective Hamiltonian, dissipative terms arise in the rigorous derivation of the effective Hamiltonian model \cite{ogamel10}.  These terms scale with $1/\omega_{nm}^-=\frac{1}{2}(1/\omega_n-1/\omega_m)$, whereas the effective Hamiltonian has terms scaling with $1/\omega_{nm}^+$, and we would like to show that these terms can be neglected in our treatment. It follows that the dissipative terms scale as either $1/\Delta_i^2$ or with the difference of deturnings, $\Delta_1-\Delta_2$, while $1/\omega_{nm}^+$ scales as either $1/\Delta_i$ or with the sum of detunings.  For an appropriate choice of the detunings, for example by making both large and of similar magnitude, we can minimize the impact of the dissipative terms which arise from time-averaging.

To study the evolution of the system we work in a dispersive regime where the internal state of the ion does not change, and it is prepared initially in one of the two levels.  All of the diagonal terms of $\bfH^{(\ell)}$ are independent of time, and act simply as phase shifts for the two normal modes which may depend on the internal state of the ion.  The terms $\bfH_{21}^{(\ell)}$ and $\bfH_{43}^{(\ell)}$ describe single-mode squeezing while the terms $\bfH_{32}^{(\ell)}$ and $\bfH_{41}^{(\ell)}$ involve two-mode squeezing between the normal modes.  The remaining terms, $\bfH_{31}^{(\ell)}$ and $\bfH_{42}^{(\ell)}$, describe the acoustic equivalent of a beam splitter between the normal modes.  It is this operation that we wish to isolate and implement in a controlled fashion.

Fortunately, each of these four types of operations oscillate at distinct frequencies and given the freedom to adjust the detunings we can make one of them resonant by eliminating its time dependence while the others will oscillate on the order of the trapping frequency $\nu$ or faster.  
Thus, for an appropriate choice of the interaction strength and detunings we can make our interaction act as a beam splitter, or potentially a two-mode squeezing operation.  The resonance conditions for these terms are as follows
\begin{align}\label{eq:res}
\delta_{31}^{(rs)}&=\nu_r-\nu_s-\Delta_1+\Delta_2\nonumber\\
\delta_{41}^{(rs)}&=\nu_r+\nu_s-\Delta_1+\Delta_2\nonumber\\
\delta_{32}^{(rs)}&=-\nu_r-\nu_s-\Delta_1+\Delta_2\nonumber\\
\delta_{42}^{(rs)}&=-\nu_r+\nu_s-\Delta_1+\Delta_2,
\end{align}
where the term is resonant when $\delta=0$.  If we choose to satisfy the first resonance condition so that the $\bfH_{31}^{(\ell)}$ mode-mixing term dominates, then the remaining three terms oscillate with frequencies: $2\nu_s,-2\nu_r,$ and $2(\nu_s-\nu_r)$.  The first two terms can be safely discarded since they are on the order of the trapping frequency, assumed to be $\sim$MHz.  The last term will still be proportional to the trapping frequency and can be neglected.
\section{Beam Splitter Derivation}\label{sec:bsplitexample}
Before looking at the full derivation and justification for neglecting off-resonant terms, we present a simple derivation for the case of a beam splitter between two normal modes $a$ and $b$.  To allow for population exchange between these modes, with frequencies $\nu_a$ and $\nu_b$, we satisfy the resonance condition
\begin{align}
\omega_{L_1}+\nu_a=\omega_{L_2}+\nu_b=\omega_{12}-\Delta,
\end{align}
to implement a two-photon process, depicted in \fig{levels}, where the difference in the laser frequencies matches the difference in the energies of the desired modes.  With this condition satisfied, the off resonant terms in the Hamiltonian of \eq{Hfull} are suppressed and the dominant contribution comes from
\begin{align}
\hat H&=-\frac{i\hbar}{2}\ket 1\bra 2\left\lbrace \Omega_1 e^{i\psi_1}\eta_1 s^{(a)}\hat a^\dagger + \Omega_2 e^{-i\psi_2}\eta_2 s^{(b)}\hat b^\dagger\right\rbrace e^{-i\Delta t}\nonumber\\&+h.c.
\end{align}
Following the procedure in \sec{effective}, where there is only one term, yields the corresponding effective Hamiltonian
\begin{align}
\hat H_{eff}&=\frac{\hbar}{4\Delta}\left[\ket 2\bra 1 (z_a\hat a+z_b\hat b),\ket 1\bra 2(z_a^*\hat a^\dagger+z_b^*\hat b^\dagger)\right]\nonumber\\
&=-\frac{\hbar}{4\Delta}\left\lbrace\Omega_1^2\eta_1^2 s^{(a)^2}\hat n_a+\Omega_2^2\eta_2s^{(b)^2}\hat n_b\right.\nonumber\\
&\left.+\Omega_1\Omega_2\eta_1\eta_2 s^{(a)}s^{(b)}e^{i(\psi_1-\psi_2)}\hat a\hat b^\dagger\right\rbrace+h.c.
\end{align}
where we define $z_a=\Omega_1 e^{i\psi_1}\eta_1 s^{(a)}$  and  $z_b=\Omega_2 e^{i\psi_2}\eta_2 s^{(b)}$.  Furthermore, have assumed the ion is initially in the state $\ket 1$ in going from the first line to the second line.  Notice that the first two terms describe an intensity-dependent phase shift while the last term is the familiar beam splitter Hamiltonian.
\section{General Analysis}\label{sec:protocol}
To derive the beam splitter operation in \sec{effective} we assumed the ability to engineer an interaction where one ion is coupled to a single normal mode.  In reality, all of the normal modes participate in the interaction as specified by \eq{Hfull}.  However, we can still write this Hamiltonian in a similar form by defining
\begin{align}\label{eq:hnprot}
\hat h_1^{(\ell,p)}&=i\hbar G_1 b_\ell^{(p)}\sqrt{\frac{\hbar}{2M\nu_p}}\hat a_p\hat \sigma_-^{(\ell)}  &&\omega_1^{(p)}= \hphantom{-}\nu_p-\Delta_1\nonumber\\
\hat h_2^{(\ell,p)}&=i\hbar G_1 b_\ell^{(p)}\sqrt{\frac{\hbar}{2M\nu_p}}\hat a_p^\dagger\hat \sigma_-^{(\ell)}  &&\omega_2^{(p)}= -\nu_p-\Delta_1\nonumber\\
\hat h_3^{(\ell,p)}&=i\hbar G_2 b_\ell^{(p)}\sqrt{\frac{\hbar}{2M\nu_p}}\hat a_{p}\hat \sigma_-^{(\ell)}  &&\omega_3^{(p)}= \hphantom{-}\nu_{p}-\Delta_2\nonumber\\
\hat h_4^{(\ell,p)}&=i\hbar G_2 b_\ell^{(p)}\sqrt{\frac{\hbar}{2M\nu_p}}\hat a_{p}^\dagger\hat \sigma_-^{(\ell)}  &&\omega_4^{(p)}= -\nu_{p}-\Delta_2.
\end{align}
With this identification we re-express the Hamiltonian $\hat H_2^{(\ell)}$ as
\begin{align}\label{Hharmonicprot}
\hat H_2^{(\ell)}&=\sum_{n,p} \hat h_n^{(\ell,p)} e^{-i\omega_n^{(p)}t}+h.c.
\end{align}
which is again in harmonic form and where the sum of any two $\omega^{(p)}$'s will be large.  We can still use the effective Hamiltonian approach provided that the detunings dominate the normal mode frequencies; a reasonable assumption.  The new $\bfH$ which characterizes the interaction will have $16N^2$ terms which we can partition into $4\times 4$ blocks containing the terms $[\hat h_m^{(\ell,p)},\hat h_n^{(\ell,p')}]$.  For distinct $p,p'$ these blocks will look the same as those from \sec{effective} defined by the terms in \eq{Heff}.  For the case where $p=p'$, the terms will describe only phase shifts and single mode squeezing operations.

\subsection{Resonance Conditions}\label{sec:resonant}
It is apparent that there are many undesired terms arising in the effective Hamiltonian.  We wish to choose the available free parameters in such a way that only two normal modes are connected, apart from phase shifts.  In particular, we wish to implement a beam splitter operation between distinct modes $r$ and $s$, with $r<s$.  In addition to the resonance conditions for the different operations in the two modes of interest, specified by \eq{res}, there will be resonance conditions for all of the other spectator modes, of the same form, which have non-zero amplitude contributing to the normal modes of interest.  The frequencies of the normal modes are specified by Eq. (13) in Ref. \cite{djames98} and are all proportional to the trapping frequency $\nu$ as well as monotonically increasing.



If we satisfy the resonance condition $\delta_{31}^{(rs)}=0$, corresponding to a beam splitter, then the resonance conditions for the spectator modes will be proportional the trapping frequency multiplied by a factor consisting of sums and differences of square-roots of eigenvalues. For relatively small chains of ions, say on the order of $\approx 10$, the smallest gap between any two eigenvalues will be such that the nearest frequency difference is $\approx 0.5\nu$.  In this case, all the undesired terms will be oscillating at $\Omega(\nu)$ and will contribute little to the evolution of the system since the integration time over which the beam splitter acts will contain many periods of these quickly oscillating terms which will average out.  As the number of ions in the chain increases, these difference terms will become smaller and it will be harder to make this same argument in general for any two modes.  However, if we are only interested in constructed operations between the lower normal modes, then we can still safely neglect the effect of the higher modes.

To recover the desired interaction Hamiltonian we simply make the identification $b_\ell^{(r)}G_1\rightarrow G_1$ and $b_\ell^{(s)}G_2\rightarrow G_2$ and choose the detunings appropriately as detailed above.  This still leaves us the freedom to choose a particular ion, the $\ell$-th one, in the chain to work with, and we can use this freedom to pick the ion that will maximize the scaling of the coupling given by $b_\ell^{(r)}b_\ell^{(s)}$.
\subsection{Limitations}\label{sec:limitations}
We have assumed the ability to cool and prepare the ions and motional modes in the ground state, and have assumed that heating is negligible on the timescale of our operations; these are the same issues trapped-ion quantum computations face in general \cite{blatt08}.  Furthermore, the interaction for a single ion with a single mode of motion is more precisely given, in the interaction picture, by \cite{wineland98}
\begin{align}\label{eq:Hactual}
\hat H_{int}^{(\ell)}&=\frac{\hbar \Omega}{2} \hat \sigma_+^{(\ell)} \exp\left\lbrace i\left[\eta\left(\hat a_\ell e^{-i\nu t}+\hat a_\ell^\dagger e^{i\nu t}\right)-t\Delta+\psi\right]\right\rbrace\nonumber\\+h.c.
\end{align}
where $\eta$ is the Lamb-Dicke parameter.  If amplitude for the ion's motion, along the axis of the trap, is much less than $k_z=k\cos\theta$, then the first-order expansion given by \eq{Hfull} is a reasonable approximation.  This condition, \break$\sqrt{\bra{\psi}k_z^2 \hat q_\ell^2\ket{\psi}} \ll 1$, is called the Lamb-Dicke regime.  In this limit, two-phonon transitions are strongly suppressed and our beam splitter approximation is valid, for more details we refer the reader to \app{beyond}.
\section{Interference Example}\label{sec:HOM}
\begin{figure}[htp]
\centering
\includegraphics[scale=.6]{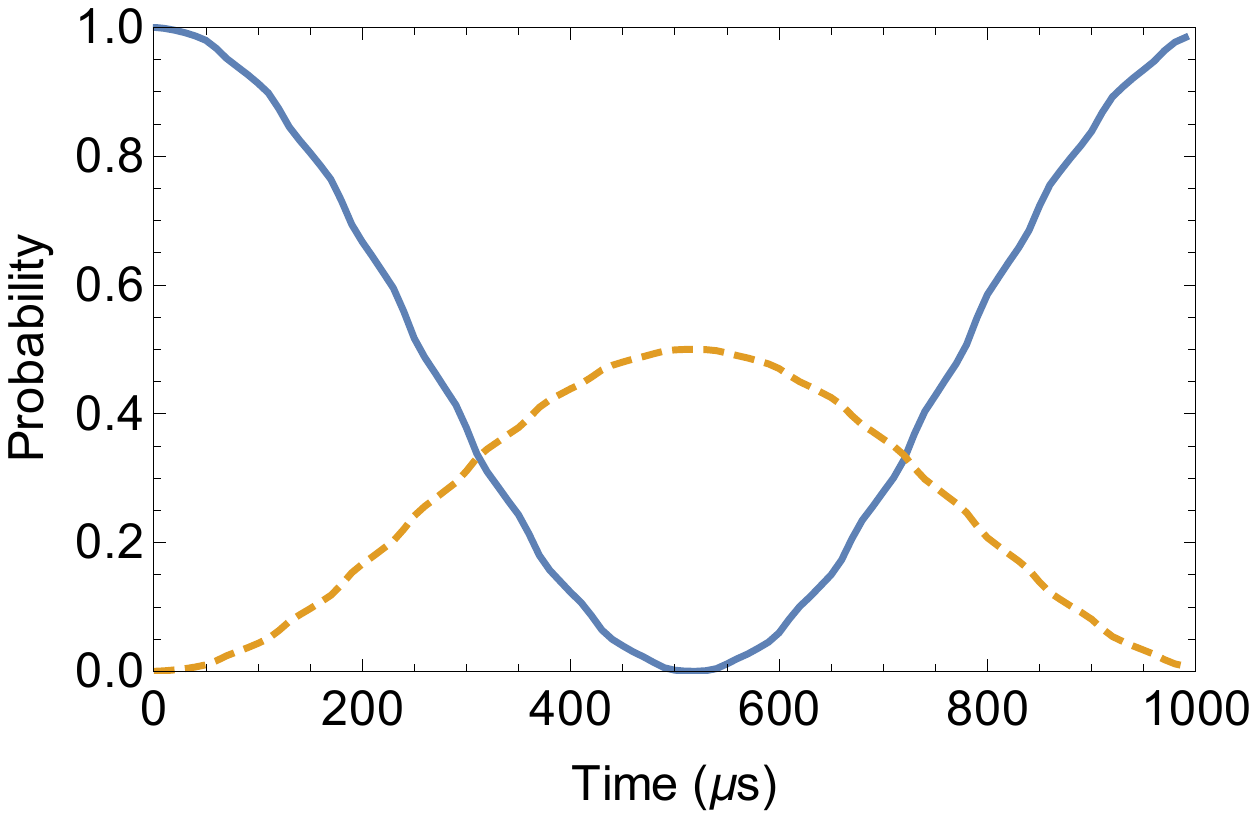}
\caption{(color online) A linear ion trap with frequency $\nu= 1$ MHz, containing two ions of atomic mass $M=6\times 10^{-26}$ kg, is initialized to have both ions in their internal ground state with one phonon in each of the two axial normal modes.  Two lasers, each characterized by a Rabi frequency $\Omega= 1$ MHz, have detunings of $\Delta_1=-1$GHz and $\Delta_2=\Delta_1-(1+\sqrt 3)\nu$ to resonantly pick out a beam splitter interaction.  The probability of detecting a phonon in each mode is given by the solid (blue) line while the probability of finding two phonons in the center-of-mass mode, $\nu_1$, is given by the dashed (orange) line.  The curve for finding two phonons in the breathing mode, $\nu_2$, is omitted as it is visually indistinguishable from the dashed line.}
\label{fig:plot1}\end{figure}

\begin{figure}[htp]
\centering
\includegraphics[scale=.6]{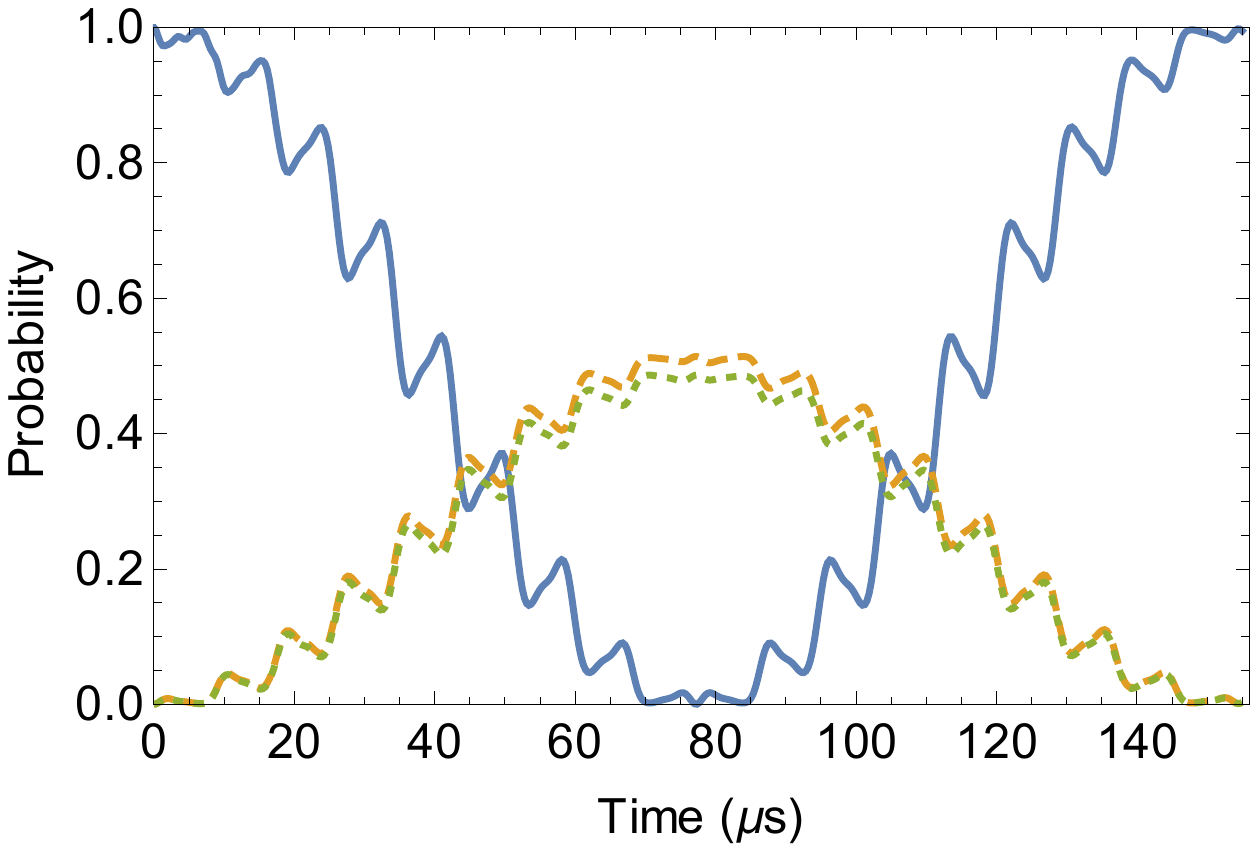}
\caption{(color online) A linear ion trap with the same specification and setup as in \fig{plot1}, however where $\Delta_1=100$MHz.  The solid line corresponds to a coincidence while the long-dashed (orange) line and short-dashed (green) line correspond to finding both phonons in the center-of-mass and breathing mode respectively.}
\label{fig:plot2}
\end{figure}

As a potential proof-of-principle example we consider a two-phonon interference experiment analogous to the Hong-Ou-Mandel experiment for photons \cite{hong87}.  Such an experiment has already been demonstrated with trapped ions and a different technique using radial modes of distinct ions instead of the normal modes in a chain \cite{toyoda15}.  We consider a chain of two ions both in their lower internal state \cite{happer69}, in our two-level approximation, and where both the center-of-mass mode, $\nu_1$, and the breathing mode, $\nu_2$, have a single excitation.  This can be achieved, for example, by cooling the joint system to its ground state \cite{heinzen90} and then resonantly exciting an ion to the excited state before using a laser tuned to the red-sideband of the desired mode.  Such preparation of non-classical motional states has been both studied and demonstrated experimentally using trapped ions \cite{meekhof96,monroe96,leibfried97}.  By choosing detunings that resonantly suppress all interaction terms except for the mode-mixing described by $\bfH_{31}^{(1)}$ we allow the system to evolve and study the the probability of one phonon remaining in each mode as a function of time.  In order to validate our results, a numerical simulation carried out within the framework of our approximations is depicted in \fig{plot1} and \fig{plot2}.

We note that a 50/50 beam splitter can be implemented by waiting a sufficiently long amount of time, the timescale of which is set by $1/\nu$.  This is evident from \fig{plot1} where we see the number of coincidences vanish for an appropriately chosen amount of time; in analogy with the Hong-Ou-Mandel dip the two phonons bunch into either mode with equal probabilities.  The dynamics on shorter timescales, and with smaller detuning is depicted in \fig{plot2} where we see qualitatively similar results but where high-frequency components affect the results more appreciably. One could experimentally implement this measurement by, for example, mapping the state of each oscillator onto the internal levels of separate ions, using red detuned lasers, and then reading these out using standard techniques \cite{itano93,leibfried96,leibfried97}.

\section{Conclusion}\label{sec:conclusions}
In the preceding sections, we have reviewed the normal mode structure for the motional states of a linear chain of trapped ions and described a method for realizing Gaussian operations between them using only two lasers of different frequencies.  In particular, we demonstrated the ability to construct a beam splitter operation for phonons, and presented considerations for the validity of the approximations considered.  Namely, we provided a justification for selecting a desired piece of the total Hamiltonian and suppressing off-resonant effects.  As a proof-of-principle example we discussed the performance of our proposal in a phonon bunching experiment analogous to the Hong-Ou-Mandel experiment for photons.

Although such operations have been discussed in the literature before \cite{klau12,duan14} their implementation is challenging and relies on engineering specific potentials or using many fast pulses.  Our approach has the benefit of requiring only modest experimental resources, namely two laser pulses of differing frequencies. By considering the converse of the conventional paradigm in trapped ion quantum information, where one only uses phonons to mediate operations on the internal degrees of freedom, we hope our work will inspire new avenues of research in harnessing the robustness of trapped ion systems to explore protocols usually reserved to the realm of photons in quantum optics.

The authors acknowledge support from NSERC and thank Rainer Blatt and Thomas Monz for useful discussions. 
\bibliographystyle{unsrt}
\bibliography{refs}

\appendix
\section{Effective Hamiltonian Terms}\label{app:Hterms}
We label the internal state of the $\ell$-th ion by the two levels $\ket 1\bra 1_\ell$ and $\ket 2\bra 2_\ell$ and define $\hat \sigma_3^{(\ell)}=\ket 2\bra 2_\ell-\ket 1\bra 1_\ell$ as the usual atomic inversion operator.  For compactness, we also define $\Sigma_{ab}=(-1)^a\nu_r+(-1)^b\nu_s-\Delta_1+\Delta_2$. Solving for the terms in the interaction Hamiltonian we find the following; note that $\bfH^{(\ell)}_{mn}=\bfH^{(\ell)\dagger}_{nm}$.
\allowdisplaybreaks\begin{align}
\bfH_{11}^{(\ell)}&=\hbar |G_1|^2\frac{\hat a_r^\dagger \hat a_r \hat \sigma_3^{(\ell)} - \ket 1\bra 1_\ell}{2M\nu_r(\nu_r-\Delta_1)}\nonumber\\
\bfH_{22}^{(\ell)}&=-\hbar |G_1|^2\frac{\hat a_r^\dagger \hat a_r \hat \sigma_3^{(\ell)} + \ket 2\bra 2_\ell}{2M\nu_r(\nu_r+\Delta_1)}\nonumber\\
\bfH_{33}^{(\ell)}&=\hbar |G_2|^2\frac{\hat a_s^\dagger \hat a_s \hat \sigma_3^{(\ell)} - \ket 1\bra 1_\ell}{2M\nu_s(\nu_s-\Delta_2)}\nonumber\\
\bfH_{44}^{(\ell)}&=-\hbar |G_2|^2\frac{\hat a_s^\dagger \hat a_s \hat \sigma_3^{(\ell)} + \ket 2\bra 2_\ell}{2M\nu_s(\nu_s+\Delta_2)}\nonumber\\
\bfH_{21}^{(\ell)}&=\hbar |G_1|^2\frac{\Delta_1(\hat a_r^\dagger)^2\hat \sigma_3^{(\ell)}}{2M(\nu_r^3-\nu_r\Delta_1^2)}e^{2i\nu_rt}\nonumber\\
\bfH_{31}^{(\ell)}&=\hbar G_1^*G_2\frac{(\nu_r+\nu_s-\Delta_1-\Delta_2)\hat a_r^\dagger\hat a_s \hat \sigma_3^{(\ell)}}{4M\sqrt{\nu_r\nu_s}(\nu_r-\Delta_1)(\nu_s-\Delta_2)}e^{it\Sigma_{01}}\nonumber\\
\bfH_{41}^{(\ell)}&=-\hbar G_1^*G_2\frac{(\nu_r-\nu_s-\Delta_1-\Delta_2)\hat a_r^\dagger\hat a_s^\dagger\hat \sigma_3^{(\ell)}}{4M\sqrt{\nu_r\nu_s}(\nu_r-\Delta_1)(\nu_s+\Delta_2)}e^{it\Sigma_{00}}\nonumber\\
\bfH_{32}^{(\ell)}&=-\hbar G_1^*G_2\frac{(-\nu_r+\nu_s-\Delta_1-\Delta_2)\hat a_r\hat a_s\hat \sigma_3^{(\ell)}}{4M\sqrt{\nu_r\nu_s}(\nu_r+\Delta_1)(\nu_s-\Delta_2)}e^{it\Sigma_{11}}\nonumber\\
\bfH_{42}^{(\ell)}&=\hbar G_1^*G_2\frac{(-\nu_r-\nu_s-\Delta_1-\Delta_2)\hat a_r\hat a_s^\dagger\hat \sigma_3^{(\ell)}}{4M\sqrt{\nu_r\nu_s}(\nu_r+\Delta_1)(\nu_s+\Delta_2)}e^{it\Sigma_{10}}\nonumber\\
\bfH_{43}^{(\ell)}&=\hbar |G_2|^2\frac{\Delta_2(\hat a_s^\dagger)^2\hat \sigma_3^{(\ell)}}{2M(\nu_s^3-\nu_s\Delta_2^2)}e^{2i\nu_st}.
\end{align}

As one can see from the above expressions, our protocol relies on a second-order effect.  One might worry that terms are missing as a result starting with a first-order Taylor expansion of the full Hamiltonian given in \eq{Hactual} and then finding effective second-order terms arising from time-ordering.  We note that terms second-order in the Lamb-Dicke parameter $\mathcal O(\eta^2)$ which would arise from further Taylor expansion would oscillate on the order of the detuning $\Delta$, and these can be safely neglected as the corrections they would impose would be $\mathcal O(\eta^4)$.
\section{Beyond the Lamb-Dicke Approximation}\label{app:beyond}
The appropriateness of the Lamb-Dicke approximation can be assessed by a more thorough treatment of our proposed method.  The true interaction Hamiltonian is given as 
\begin{align}\label{eq:Htrue}
\hat H_{int}^{(\ell)}&=\frac{\hbar \Omega}{2} \hat \sigma_+^{(\ell)} \exp\left\lbrace i\left[\eta\left(\hat a_\ell e^{-i\nu t}+\hat a_\ell^\dagger e^{i\nu t}\right)-t\Delta+\psi\right]\right\rbrace\nonumber\\+h.c.
\end{align}
Using an identity \cite{cahill69} we can express $\Omega_{n'n}=\Omega\bra{n'}e^{i\eta(a+a^\dag)}\ket n$ as
\begin{align}
\Omega_{n'n}&=\Omega e^{-\eta^2/2} \sqrt{\frac{n_<!}{n_>!}} \eta^{|n'-n|} L_{n_<}^{|n'-n|}(\eta^2),
\end{align}
where $n_>$ ($n_<$) is the greater (lesser) of $n'$ and $n$, and $L_n^\alpha$ is the generalized Laguerre polynomial
\begin{align}
L_n^\alpha(x)=\sum_{m=0}^n(-1)^m\binom{n+\alpha}{n-m}\frac{x^m}{m!}.
\end{align}
By noting that
\begin{align}
&\Omega\bra{n'}\exp{\left[i\eta\left(\hat ae^{-i\nu t}+\hat a^\dag e^{i\nu t}\right)\right]}\ket n\nonumber\\
&=\Omega\bra{n'} e^{i \nu \hat a^\dag \hat a t} \exp{\left[i\eta\left(\hat a+\hat a^\dag\right)\right]} e^{-i \nu \hat a^\dag \hat a t}\ket n\nonumber\\
&=\Omega\bra{n'} \exp{\left[i\eta\left(\hat a+\hat a^\dag\right)\right]} \ket n e^{i\nu(n'-n) t}\nonumber\\
&=\Omega_{n'n}e^{i\nu(n'-n)t},
\end{align}
we can now write the Hamiltonian as
\begin{align}
\hat H_{int}^{(\ell)}&=\frac{\hbar}{2} \hat \sigma_+^{(\ell)} \sum_{n',n} \Omega_{n'n} \ket{n'}\bra n e^{-i(\Delta t-\psi)}e^{i\nu(n'-n)t}+h.c.
\end{align}
where we denote $\Gamma=n'-n$ to emphasize that terms with fixed $\Gamma$ have the same frequency.  Motivated by this we can express the interaction as
\begin{align}
\hat H_{int}^{(\ell)}&=\frac{\hbar}{2} \hat \sigma_+^{(\ell)} \sum_{n,\Gamma} \Omega_{n+\Gamma,n} \ket{n+\Gamma}\bra n e^{-i(\Delta t-\psi)}e^{i\nu\Gamma t}+h.c.\nonumber\\
&=\sum_\Gamma \hat h_\Gamma e^{-i(\Delta-\nu\Gamma)t}+h.c.
\end{align}
where
\begin{align}\label{eq:hbeyond}
\hat h_\Gamma&=\frac{\hbar}{2} e^{i\phi} \hat \sigma_+^{(\ell)} \sum_n \Omega_{n+\Gamma,n}\ket{n+\Gamma}\bra n.
\end{align}
Provided the detuning $\Delta$ dominates this frequency over some appropriate range of $\Gamma$, such as if we had an upper bound on the number of phonons, then we can again appeal to the effective Hamiltonian approach.  We can then compare these results to the results obtained in the Lamb-Dicke approximation.  

For example, we compare the term containing $\eta \hat \sigma_+^{(\ell)}\hat a$ obtained in the Lamb-Dicke approximation to the corresponding term in \eq{hbeyond}.  The former will have the matrix element $\eta \bra{n'}\hat a \ket n=\eta \sqrt{n}\delta_{n',n-1}$ whereas the latter will have
\begin{align}
\bra{n'}\left(\sum_m \Omega_{m+\Gamma,m}\ket{m+\Gamma}\bra m\right)\ket n&=\Omega_{n+\Gamma,n}\delta_{n',n+\Gamma}
\end{align}
For the case of $\Gamma=-1$ we find that $\lim_{\eta\rightarrow 0}\Omega_{n-1,n}\approx \eta \sqrt{n}$ as desired and all other transitions are suppressed by a factor of $\eta^{|\Gamma|}$.  Deviations from the first-order approximation can then be characterized by the how these matrix elements differ from the creation and annihilation operators assumed in the more-simple treatment.

One can account for the transverse modes by adding additional terms to the exponential in \eq{Htrue}, these terms will give rise to similar terms $\hat h_\Gamma$ in the effective Hamiltonian picture, i.e., phase shifts, beam splitters, and squeezing operations.  As presented in \sec{protocol} these motional modes can be expressed in terms of the normal modes of the chain.  
\end{document}